\documentclass[prd,aps,twocolumn,superscriptaddress,preprintnumbers,nofootinbib,showpacs]{revtex4}
\usepackage{graphicx}
\usepackage{exscale}
\usepackage[intlimits]{amsmath}
\usepackage{amsfonts}
\usepackage{amssymb,amscd}
\usepackage{epsfig}
\usepackage{pstricks}

\newcounter{commentdepth}
\newcommand{\be}{\begin{equation}}
\newcommand{\ee}{\end{equation}}

\newcommand{\beqa}{\begin{eqnarray}}
\newcommand{\eeqa}{\end{eqnarray}}
\newcommand{\beq}{\begin{equation}}
\newcommand{\eeq}{\end{equation}}

\newcommand{\p}{\partial}

\newcommand{\reftitle}[1]{}

\begin{document}
%%\preprint{\parbox[t]{\textwidth}
%%{\small May 26, 2008 \hspace*{5mm} \#\#\# \hfill 
%%ArXiv:0805.3804 [hep-ph]}}

\title{Some exact infrared properties of gluon and ghost propagators\\ and long-range force in QCD}

%\author{Klaus~Lichtenegger}%
%\affiliation{Karl-Franzens-Universit\"at Graz, 8010 Graz, Austria}%

\author{Daniel~Zwanziger}
\affiliation{Physics Department, New York University, New York, NY 10003, USA}

\begin{abstract}
\noindent We derive some exact relations in Landau gauge that follow from a cut-off at the Gribov horizon which is then implemented by a local, renormalizable action involving auxiliary bose and fermi ghosts.  The fermi ghost propagator is more singular than $1/k^2$ at $k = 0$, and the relation $\alpha_D + 2 \alpha_G = (D - 4)/2$ holds between the infrared critical exponents of the gluon and ghost propagators $D(k)$ and $G(k)$ in $D$ Euclidean dimensions.  Finally, in $D$ Euclidean dimensions, there is a long-range force, transmitted by the propagator of the auxiliary bose ghost that corresponds to a linearly rising potential with tensor coupling to colored quarks that is proportional to the renormalization-group invariant $g^2 D(k) G^2(k)$.  A comparison with numerical results is discussed.
\end{abstract}


\maketitle

\section{Introduction}

	The  present study is motivated by recent numerical investigations on very large lattices \cite{Cucchieri:2008.12, Cucchieri:2008.04, Ilgenfritz:2009, Oliveira:2008, vonSmekal2007}, in which it is reported that, in $D = 3$ and 4 Euclidean dimensions in minimal  Landau gauge,\footnote{Here the gauge is fixed at a relative minimum of the minimizing functional (\ref{minimizing}).} the gluon propagator~$D(k)$ approaches a finite value at~$k = 0$, while the ghost propagator $G(k)$ behaves like~$1/k^2$.  These properties are in contradiction with a scenario developed by Gribov \cite{Gribov:1977wm} and elaborated by the author \cite{Zwanziger:1989mf, Zwanziger:1993} according to which the gluon propagator~$D(k)$ is suppressed and even vanishing at $k = 0$, while the ghost propagator $G(k)$ is enhanced at $k =0$ compared to $1/k^2$.  On the other hand numerical results in $D = 2$ Euclidean dimensions are consistent with this scenario \cite{Maas:2007, Cucchieri:2007b}, and moreover in a detailed numerical study of Landau gauge in $D = 3$ Euclidean dimensions, it was found that if one gauge fixes to the extent practicable to an absolute --- instead of merely a relative --- minimum of the minimizing functional, then results are also not inconsistent with the Gribov scenario~\cite{Maas:2008}.  It was also observed that in~4 dimensions it would be prohibitively costly to seek an absolute minimum on lattices of sufficiently large size~\cite{Maas:2008}.
	
	In this unclear situation, it seemed worthwhile to see what exact properties could be derived by analytic methods, if the functional integral is cut-off at the Gribov horizon.  We start by reviewing previous results which are scattered in different articles.  We then obtain as exact infrared properties in~$D$ Euclidean dimensions (i) that the ghost propagator~$G(k)$ is more singular than $1/k^2$, (ii) that the equation~\cite{Alkofer:2000, Zwanziger:2001, vonSmekal:2002},
\beq
\label{exponentrelation}
\alpha_D + 2 \alpha_G = (D - 4)/2,
\eeq
holds, where the gluon and ghost infrared exponents $\alpha_D$ and $\alpha_G$ are defined by, 
\beq
\label{infracrit}
D(k) \sim { 1 \over (k^2)^{1 +  \alpha_D} }; \ \ \ \ \ \ G(k) \sim { 1 \over (k^2)^{1 +  \alpha_G} },
\eeq
and (iii) that
there is a long-range linearly rising renormalization-group invariant potential that has tensor coupling to colored quarks.

\section{Absolute Landau gauge}

	The absolute Landau gauge is defined by gauge transforming to the absolute minimum of the functional \cite{Nakajima:1978}
\beq
\label{minimizing}
F_A(g) = \| ^g A \|^2,
\eeq	
with respect to all gauge transformations $g$, where $^gA =  g^{-1}Ag + g^{-1}\p g$ is the gauge transform of $A$, and $\| A \|$ is the Hilbert norm of $A$.  From stationarity at a relative or absolute minimum, it follows that $A$  is transverse, $\p \cdot A = 0$, so the fundamental modular region $\Lambda$, which is defined to be the set of absolute minima of the minimizing functional $F_A(g)$ on each gauge orbit, satisfies
\beq
\Lambda \equiv \{A: \p \cdot A = 0, \| A\| \leq \|^gA\| \},
\eeq
for all gauge transformations $g$.

	The set of relative minima, called the Gribov region~$\Omega$, is larger than the set of absolute minima, $\Lambda \subset \Omega$, and may be characterized as the region, 
where $A$ is transverse and the Faddeev-Popov operator is non-negative, $M(A) \geq 0$, which means all its eigenvalues are non-negative,
\beq
\label{defineOmega}
\Omega\equiv \{ A: \p \cdot A = 0, \ M(A) \geq 0 \}.
\eeq
The Faddeev-Popov operator is defined by
\beq
M^{ac}(A) \equiv - D_\mu^{ac} \p_\mu  =
- \p^2 \delta^{ac} - g f^{abc} A_\mu^b \p_\mu,
\eeq
where $D_\mu^{ac}$ is the gauge-covariant derivative and $a, b, c$ are color indices, with $a = 1, ... \ N$ for SU(N).  The positivity property holds because at a relative or absolute minimum the matrix of second derivatives is positive which, for the minimizing functional (\ref{minimizing}), is the operator $M(A)$.

Recall that the fundamental modular region $\Lambda$ enjoys the following simple properties \cite{Semenov:1982}. (i) The configuration $A = 0$ is an interior point of $\Lambda$.  (ii) $\Lambda$ is bounded in every direction, which means that if $A \in \Lambda$ then for some sufficiently large positive number $\lambda > 1$ the configuration $\lambda A \notin \Lambda$.  (iii)  $\Lambda$ is convex, which means that if $A_1 \in \Lambda$ and $A_2 \in \Lambda$, than for positive numbers $\alpha > 0$ and $\beta > 0$ with $\alpha + \beta = 1$, then the linear combination, $\alpha A_1 + \beta A_2 \in \Lambda$, lies in $\Lambda$.  The Gribov region $\Omega$ enjoys the same three properties \cite{Zwanziger:1982}.

\section{Cut-off functional integral}
		
	In the absolute  Landau gauge the partition function is expressed as an integral over $\Lambda$, 
\beq
\label{partitionLambda}
Z_\Lambda  =  N \int_\Lambda dA \ \delta(\p \cdot A)\  \exp[-S(A)] ,
\eeq	
Here
\beq
\label{weightLambda}
S(A) \equiv  - \ln\det(M(A)) + S_{\rm YM}(A)) - \ln\det(\not \! \! D + m_q),
\eeq
is an effective action that includes the Faddeev-Popov and quark determinants,  and $S_{YM}$ is  the Euclidean Yang-Mills action.  The
Faddeev-Popov determinant, $\det(M(A))$, is positive for $A$ in $\Lambda$ because $\Lambda \subset \Omega$, and $M(A)$ is a positive operator in $\Omega$.

	We do not have an explicit analytic expression for the boundary of the fundamental modular region $\Lambda$.  However we know that it is contained in the Gribov region~$\Omega$ and that part of their boundary is common.  This part includes Abelian configurations and, on the lattice, center-vortex configurations which, in some scenarios, are the dominant configurations \cite{Greensite:2004}.  It has been argued that for purposes of the functional integral, the fundamental modular region $\Lambda$ and the Gribov region $\Omega$ are equivalent \cite{Zwanziger:1994, Zwanziger:2003a}.   In the present article we take this as a working hypothesis, and calculate Euclidean averages by integrating over $\Omega$ instead of $\Lambda$, with partition function
\beq
\label{partitionOmega}
Z_\Omega  =  N \int_\Omega dA \ \delta(\p \cdot A)\  \exp[-S(A)] .
\eeq
We shall deduce some exact properties of the gluon and ghost propagators, and discuss the validity of the results obtained in the concluding section.

	When comparing with numerical determinations of the gluon and ghost propagators, it should be kept in mind that the lattice analog of integrating over $\Omega$ is not the same as numerical gauge fixing to a relative minimum by a particular algorithm.  For although such an algorithm does give a gauge with a probability distribution that lies within $\Omega$, this gauge is, in principle, algorithm-dependent.  This distinction may be the origin of the discrepancy noted in the Introduction between the properties we shall obtain here and numerical results obtained by fixing to a relative minimum.  
	
	The advantage of integrating over the Gribov region~$\Omega$ is that it is explicitly described as the set of transverse configurations $A$ for which the ``horizon function" $H(A)$ has the upper bound
\beq
\Omega = \{A: \p \cdot A = 0, \ H(A) \leq qV \},
\eeq
where $V$is the Euclidean volume, and
\beq
q \equiv (N^2 - 1)D,
\eeq
is the number of components of the gauge field $A_\mu^b$.  Points $A$ on the boundary $\p \Omega$ of the Gribov region satisfy $H(A) = q V$.  The horizon function is given explicitly by~\cite{Zwanziger:1989mf},  
\beq
\label{horizfunct}
H = g^2(A, M^{-1} A) = \int d^Dx \ h(x),
\eeq
where $h(x)$ is the density
\beq
\label{horizdens}
h(x) = g^2 f^{abc} A_\mu^b(x) \int d^Dy \  (M^{-1})^{ce}(x, y; A) f^{ade} A_\mu^d(y).
\eeq
This expression for $H(A)$ resembles the formula for the $T$-matrix in potential scattering at zero energy, $T = V - V H^{-1}V$, where $H$ is the Hamiltonian.  It was obtained by a quantum mechanical-type calculation that consists in summing the perturbation series for the lowest non-trivial eigenvalue of $M(A)$.  Note that $H(A) > 0$ is positive for $A$ in $\Omega$.  The cut-off is non-perturbative because perturbatively we have $H(A) = O(g^2)$, while the cut-off occurs at $H(A) = O(1)$.

	We write the partition function $Z_\Omega$ as	
\beq
\label{parttheta}
Z_\Omega = N \int dA \ \delta(\p \cdot A) \ \theta(qV - H) \ \exp[ - S(A) ],
\eeq	
where the restriction to $\Omega$ is effected by the $\theta$-function.

\section{Boltzmann-type distribution}

	In this section we shall convert the functional integral~$Z_\Omega$ with a cut-off at the boundary of~$\Omega$ into a Boltzmann-type distribution.

	For this purpose we represent the $\theta$-function by its fourier decomposition,
\beq
\label{represent}
\theta(qV - H) = \int { d \omega \over 2 \pi i } \ { \exp[i \omega(qV - H)] \over \omega - i \epsilon }.
\eeq
and perform the $A$-integration first, so
\beq
\label{Zascontour}
Z_\Omega =  \int { d \omega \over 2 \pi } \ { \widetilde{Z}( i \omega) \over \epsilon + i \omega }.
\eeq
Here
\beq
\label{extended}
\widetilde Z(\gamma) \equiv N \int dA \ \delta(\p \cdot A) \ \exp[ - S + \gamma(q V - H) ],
\eeq 
is a Boltzmann-type partition function that depends on a thermodynamic parameter $\gamma$ that is the analog of $\beta = 1/kT$, and that may be complex.  The denominator in~(\ref{Zascontour}) is analytic in the lower half $\omega$-plane defined by $\omega \to \omega - i \gamma$, or $i \omega \to \gamma + i \omega$, where $\gamma > 0$ is positive.  The partition function $\widetilde Z(\gamma + i \omega)$ is also analytic in $z = \gamma + i \omega$ for  $\gamma > 0$.  Indeed $\exp[i \omega (qV - H)]$ is an entire analytic function of $\omega$, and moreover in the lower half plane the factor $\exp[ - (\gamma + i \omega)H(A)]$ only improves convergence for the $A$-integration because  $H(A) > 0$.  Thus we may deform the contour of integration in~(\ref{Zascontour}) into the lower half-plane, 
\beq
\label{omegaintegral}
Z_\Omega = \int_{- \infty}^{+ \infty} d\omega \ \exp[\widetilde W(\gamma + i \omega) - \ln (\gamma + i \omega)],
\eeq
where the free energy is defined by 
\beq
\widetilde W(\gamma + i \omega) \equiv \ln [\widetilde Z(\gamma + i \omega)].
\eeq
It is a bulk quantity, 
\beq
\widetilde W(\gamma) = V \widetilde w(\gamma) + O(1).
\eeq
Here $V$ is of the order of the total number of degees of freedom which, with a lattice cut-off, is of the order of the volume of the lattice in lattice units, $V/a^D$.  We are interested in the limit $V/a^D \to \infty$, which we write as~$V \to \infty$.

	For large $V$, the exponent $\widetilde W(\gamma) = O(V)$ in (\ref{omegaintegral}) is large, so the contour integration over~$\omega$ may be evaluated by the saddle-point method.  To leading order in $V$ in the exponent, the term $\ln(\gamma + i \omega)$, which is of order 1, may be neglected, and the saddle point,~$\gamma_s$, if it exists, is the solution of 
\beq
\label{saddlepoint}
{ \p  \widetilde W(\gamma_s) \over \p \gamma } = 0.
\eeq
This saddle-point condition is a gap equation that relates~$\gamma_s$ to the mass scale $\Lambda_{QCD}$ \cite{Zwanziger:1989mf}.  

	In the neighborhood of the saddle-point, $\widetilde W$ has the expansion
\beq
\widetilde W(\gamma_s + i \omega) = \widetilde W(\gamma_s) - { \p^2 \widetilde W(\gamma_s) \over \p \gamma^2 } \ { \omega^2 \over 2 },
\eeq
where, by (\ref{extended}),
\beq
\label{variance}
{ \p^2 \widetilde W(\gamma_s) \over \p \gamma^2 } = \langle H^2 \rangle_{\gamma_s} - \langle H \rangle_{\gamma_s}^2 > 0.
\eeq
The subscript refers to an expectation-value calculated in the Boltzmann-type ensemble (\ref{extended}) namely,
\beqa
\label{Boltzmanntype}
\langle X \rangle_{\gamma} & \equiv & \widetilde Z^{-1}(\gamma) 
\\ \nonumber
&& \times \ N \int dA \ \delta(\p \cdot A) \ \exp[ - S + \gamma(q V - H) ] \ X.
\eeqa
The quantity on the right of (\ref{variance})
is the variance of the horizon function $H$.  It is positive and of order $V$, so the width of the peak in the $\omega$-integration is of order $V^{-1/2}$.  Thus the saddle-point approximation becomes exact in the limit $V \to \infty$, and we obtain the basic relation 
\beq
Z_\Omega =  \widetilde Z(\gamma_s).
\eeq
This equality represents symbolically the equality of expectation values calculated with the cut-off distribution $Z_\Omega$ and the Boltzmann-type distribution $\widetilde Z(\gamma_s)$ as in (\ref{Boltzmanntype}).

\subsection{Remarks} (i)~We would have gotten the same result if the $\theta$-function, $\theta(qV - H)$ in~(\ref{parttheta}), were replaced by the $\delta$-function $\delta(qV - H)$.  Indeed we have seen that the term $\ln(\gamma + i \omega)$ in~(\ref{omegaintegral}) may be neglected to leading order in~$V$.  This term comes from the denominator $\omega - i \epsilon$ in~(\ref{represent}), so we have in fact replaced the denominator by 1.  This is the same as replacing $\theta(qV - H)$ by the $\delta$-function,
\beq
\delta(qV - H) = \int { d\omega \over 2 \pi } \exp[ i \omega(fV - H)],
\eeq
whose fourier decomposition has no denominator.  (ii)~Since $\delta(qV - H)$ is entirely concentrated on the boundary $H = qV$ of the fundamental modular region, it follows that at large $V$ the probability gets concentrated on the boundary.  (iii)~The microcanonical ensemble $\delta(qV - H)$, and the equivalent cut-off ensemble $\theta(qV - H)$, are also equivalent to a canonical or Boltzmann-type ensemble (\ref{extended}), where the parameter $\gamma$ is the analog of $\beta = {1 \over kT}$.\footnote{Here we use the terms ``microcanonical" and ``canonical" by analogy with the corresponding distributions in statistical mechanics to which they are mathematically, but not physically identical.}   For, by (\ref{extended}), the saddle-point condition, ${\p \widetilde W(\gamma ) \over \p \gamma } = 0$, with $\widetilde W(\gamma) = \ln \widetilde Z(\gamma)$, is equivalent to $\langle qV - H \rangle_{\gamma} = 0$, or
\beq
\label{horizoncond1}
\langle H \rangle_{\gamma} = qV,
\eeq
where the subscript is defined in (\ref{Boltzmanntype}).  Thus the ``horizon condition" $H = qV$, which would be imposed by $\delta(qV - H)$ is satisfied in the mean in the Boltzmann-type distribution.  Moreover, since $H(A)$ is a bulk quantity, its fluctuations are of relative order $1/V^{1/2}$ which is negligible.  Thus the saddle-point condition is the horizon condition.

\section{Local renormallizable action}

	We wish to represent the partition function $\widetilde Z(\gamma)$, which contains the non-local action $S + \gamma(H - fV)$ as a functional integral with a local action.  We use the Faddeev-Popov identity as usual to rewrite $\delta(\p \cdot A) \det M(A)$ as an integral over additional fields, with the local Faddeev-Popov action
\beqa
\int dA \  \delta(\p \cdot A) \ \det M(A) \ \exp(-S_{YM}) 
\nonumber  \\
= \int dA db dc d\bar c \ \exp(-S_{FP}),
\eeqa	
where $S_{FP} = \int d^Dx \ {\cal L}_{FP}$ is the local Faddev-Popov action in Landau gauge
\beq
{\cal L}_{FP} = {1 \over 4} \ F_{\mu\nu}^2 + i \p_\mu b A_\mu - \p_\mu \bar c D_\mu c, 
\eeq	
and
\beq
F_{\mu \nu} = \p_\mu A_\nu - \p_\nu A_\mu + g A_\mu \times A_\nu,
\eeq
where $(A_\mu \times A_\nu)^a \equiv f^{abc} A_\mu^b A_\nu^c$.  The Lagrange-multiplier field $b$ imposes the Landau gauge condition $\p \cdot A = 0$.  Without loss of generality we ignore the quark action which plays no role in the discussion.  

	As a second step, we similarly eliminate the non-local term $\gamma(H - qV)$ in the action (\ref{extended}) by an integral over auxiliary fields and a local action.  This is accomplished by the identity \cite{Zwanziger:1989mf, Zwanziger:1993},
\beq
\label{integrateout}
\exp[- \gamma(H - qV)] = \int d\phi d\bar\phi d\omega d\bar\omega \ \exp(-S_{aux} - S_\gamma),
\eeq
that is easily verified by integrating out the auxiliary fields by Gaussian integration.  They consist of a quartet of auxililary bose and fermi ghosts, $\phi_\mu^{ab}$ and $\omega_\mu^{ab}$, and corresponding anti-ghosts $\bar\phi_\mu^{ab}$ and $\bar\omega_\mu^{ab}$ that carry a Lorentz index and a pair of color indices.  The term on the left, $\gamma H = \gamma g^2 (A, M^{-1}A)$, is obtained by completing the square in the exponent in the Gaussian integration of the bose ghosts $\phi$ and $\bar \phi$.  This produces an unwanted $1/\det^q(M)$ in the denominator which is then cancelled by $\det^q(M)$ in the numerator that comes from the integrating out the fermi ghosts $\omega$ and $\bar \omega$.  The total local action is given by
\beq
\widetilde S = \int d^Dx \ {\cal L} 
\eeq
where
\beq
\label{Lph}
{\cal L} = {\cal L}_{FP} + {\cal L}_{aux} + {\cal L}_\gamma,
\eeq   
and
\beqa
\label{auxact}
{\cal L}_{aux} & = &
  \p_\lambda \bar\phi_\mu^{ab} (D_\lambda \phi_\mu)^{ab}
  \\ \nonumber  
&&   - \p_\lambda \bar\omega_\mu^{ab} 
[ \ (D_\lambda \omega_\mu)^{ab} + (g D_\lambda c \times \phi_\mu)^{ab} \ ].
\eeqa 
The gauge-covariant derivative and the Lie commutator act on the first color index only, while the second color index is mute, thus 
$(D_\lambda \phi_\mu)^{ab} = \p_\lambda \phi_\mu^{ab} + g (A_\lambda \times \phi_\mu)^{ab}$
where $(A_\lambda \times \phi_\mu)^{ab} \equiv f^{acd}A_\lambda^c \phi_\mu^{db}$.  The last term involves the thermodynamic parameter $\gamma$,
\beq
\label{Lgamma}
{\cal L}_\gamma =  
 \gamma^{1/2} \ 
[ \  D_\lambda(\phi_\lambda - \bar\phi_\lambda)
-  g(D_\lambda c \times \bar\omega_\lambda) \ ]^{aa}
 - \gamma q.
\eeq 
If the term ${\cal L}_\gamma$ were absent, the integral over the bose and fermi ghosts would produce cancelling factors of $\det M$, and the action would be strictly equivalent to the Faddeev-Popov action.   

	All terms except ${\cal L}_\gamma$ are of dimension 4 and respect a BRST symmetry transformation that acts on the Faddeev-Popov fields according to
\beqa
\label{BRST1}
s A_\mu & = & D_\mu c; \ \ \ \ \ \ \  \ \ \ \ \  sc = - (g/2) c \times c 
\nonumber  \\
s \bar c & = & ib; \ \ \ \ \ \ \ \  \ \ \ \ \ \ \    s b = 0, 
\eeqa	
and acts trivially on the auxiliary ghosts,
\beqa
\label{BRST2}
s \phi_\mu^{ab} & = & \omega_\mu^{ab}; 
\ \ \ \ \ \ \ \  s \omega_\mu^{ab} = 0
\nonumber  \\
s \bar\omega_\mu^{ab} & = & \bar\phi_\mu^{ab}; 
\ \ \ \ \ \ \ \  s \bar\phi_\mu^{ab} = 0.
\eeqa	
It is nil-potent, $s^2 = 0$.  The term
\beqa
\label{auxacta}
{\cal L}_{aux} & = &  s \ 
\p_\lambda \bar\omega_\mu^{ab} (D_\lambda \phi_\mu)^{ab}
\nonumber  \\
& = & \p_\lambda \bar\phi_\mu^{ab} (D_\lambda \phi_\mu)^{ab}
  \\   \nonumber
&& - \p_\lambda \bar\omega_\mu^{ab} 
[ \ (D_\lambda \omega_\mu)^{ab} + (g D_\lambda c \times \phi_\mu)^{ab} \ ].  
\eeqa 
is $s$-exact.	
	
	This symmetry is softly broken by the term in ${\cal L}_\gamma$ that is of dimension 2, so renormalizability is preserved \cite{Zwanziger:1993, Schaden:1994, Sorella:2005}.  [The terms in ${\cal L}_{aux}$ and ${\cal L}_\gamma$ that contain the Faddeev-Popov ghost $c$ and the auxiliary ghost $\bar\omega$ play no dynamical role because there is no corresponding term that involves $\bar c$ and $\omega$, so these terms may be eliminated by a suitable translation of $\omega$.]
	
	The local, renormalizable action $\widetilde S(\gamma)$ is an ``extended" action in the sense that it depends on the arbitrary parameter $\gamma$.  It corresponds to a gauge theory only when the saddle-point condition (\ref{horizoncond1}) is satisfied.

\section{Infrared limit of fermi ghost propagator}

	In this section we show that the horizon condition (\ref{horizoncond1}) is equivalent to the condition that the fermi-ghost propagator $G(k)$ be more singular than $1/k^2$ at $k = 0$,
\beq
\label{singghost}
\lim_{k \to 0} [k^2 G(k)]^{-1} = 0.
\eeq	
Here
\beq
\delta^{ab} G(x-y) = \langle c^a(x) \bar c^b(y) \rangle = \langle (M^{-1})^{ab}(x, y; A) \rangle
\eeq
is the fermi-ghost propagator, and $G(k)$ is its fourier transform.  This form of the horizon condition makes manifest that it is multiplicatively renormalizable.  This result was obtained previously \cite{Zwanziger:1993, Zwanziger:1994}.  We give here a simpler, more direct proof.
 
		As a first step we develop a useful expression for the horizon or saddle-point condition (\ref{horizoncond1}) in terms of local fields.  By translation invariance and (\ref{horizfunct}), the horizon condition  may be written
\beq
\label{densecondition}
\langle h(0)  \rangle = q,
\eeq		
where the expectation value is calculated in the ensemble defined by the action $\widetilde S$.  The kernel of the inverse Faddeev-Popov operator is the expectation value of the fermi-ghost correlator at fixed $A$,
\beq
\label{kernel}
\langle c^a(x) \bar c^b(y) \rangle_A =  (M^{-1})^{ab}(x, y; A).
\eeq
This, and expression (\ref{horizdens}) for $h(x)$, yields
\beq
\label{meanh0}
\langle h(0)  \rangle =
g^2\langle \ f^{abd} A_\mu^b(0) c^d(0) \int d^Dy \  f^{ace} A_\mu^c(y) \bar c^e(y) \ \rangle.
\eeq
In terms of the 4-point function 
\beq
\label{4pointfn}
\Delta_{\mu \ \nu}^{b d c e }(x, z, y, w) = \langle  A_\mu^b(x) \ c^d(z)  \  A_\nu^c(y) \ \bar c^e(w) \rangle
\eeq
this reads
\beq
\label{4pthorizon}
\langle h(0)  \rangle =
g^2 f^{abd} \ f^{ace} \int d^Dy \ \Delta_{\mu \ \nu}^{b d c e }(0, 0, y, y).
\eeq
The integral over $y$ projects onto the zero-momentum component, and we have
\beq
\label{4pthorizona}
\langle h(0)  \rangle =
g^2 f^{abd} \ f^{ace} \Delta_{\mu \ \nu}^{b d c e }(k)|_{k = 0}.
\eeq
where
\beq
\Delta_{\mu \ \nu}^{b d c e }(k)  \equiv \int d^Dx \ \exp( i k \cdot y) \ \Delta_{\mu \ \nu}^{b d c e }(0, 0, y, y)
\eeq
is a fourier transform.  The 4-point function (\ref{4pointfn}) has the cluster decomposition into connected parts, 
\beqa
\Delta_{\mu \ \nu}^{b d c e }(x, z, y, w) = \langle  A_\mu^b(x) \  A_\nu^c(y) \rangle \langle c^d(z)  \   \bar c^e(w) \rangle 
\nonumber  \\
+ {\Delta_{con,}}_{\mu \ \nu}^{b d c e }(x, z, y, w),
\eeqa
which gives
\beq
\Delta_{\mu \ \nu}^{b d c e }(k) = (D_{\mu \nu}^{b c} \ G^{de})(k) + {\Delta_{con,}}_{\mu \ \nu}^{b d c e }(k),
\eeq
where
\beq
D_{\mu \nu}^{bc}(x-y) = \langle A_\mu^b(x) A_\nu^c(y) \rangle
\eeq
is the gluon propagator.  Thus we obtain finally
\beq
\label{4pthorizonc}
\langle h(0) \rangle =  g^2 f^{abd} \ f^{ace}[  (D_{\mu \nu}^{b c} \ G^{de}) + {\Delta_{con,}}_{\mu \ \nu}^{b d c e } \ ](k)|_{k = 0},
\eeq
and the horizon condition $q =  \langle h(0) \rangle$ is expressed by
\beq
q -  g^2 f^{abd} \ f^{ace}[  (D_{\mu \nu}^{b c} \ G^{de}) + {\Delta_{con,}}_{\mu \ \nu}^{b d c e } \ ](k)|_{k = 0} = 0.
\eeq

	We now compare this form of the horizon condition with the Dyson-Schwinger equation (DSE) for the ghost propagator $G$, which we write symbolically  to avoid a profusion of indices, 
\beqa
\label{ghostDSE}
G^{-1}(k) & = & k^2 - g f k \ \sum {\rm ( loop)}(k)
\nonumber  \\
& = & k^2 - g f k \ \sum_\Phi G_{c \bar c} D_{A \Phi} \Gamma_{c, \Phi, \bar c}(k).
\eeqa
This expression is summed over loops namely over fields~$\Phi$ that have non-zero propagators with $A$ namely $\Phi = A, \phi$ or $\bar \phi$.  Here $k^2$ is the tree-level term, $g f k$ is the tree level vertex, where $f$ represents $f^{abc}$ which is contracted as in (\ref{4pthorizonc}).  The external ghost momentum $k$ factors out of the loop integral because of the transversality of the gluon propagator in Landau gauge, $(k - p)_\mu D_{\mu \nu}(p) = k_\mu D_{\mu \nu}(p)$.

	The DSE for the 3-point vertex reads
\beq
\Gamma_{c, \Phi, \bar c} = k f g + \sum {\rm (loop)} \ k f g
\eeq
where the external anti-ghost momentum $k$ again factors out of the loop.  When this is substituted into the ghost DSE, both the ghost and anti-ghost momenta factor out, and it reads
\beq
\label{PiDSE}
G^{-1}(k) = k_\mu \Pi_{\mu \nu}(k) k_\nu
\eeq
where
\beq
\label{pimunu}
\Pi_{\mu \nu}(k) = \delta_{\mu \nu} - g f \ [(D_{\mu \nu} G)  
+  \ \Delta'_{con, \mu \nu}] (k) \ f g.
\eeq
Upon examining the sum of the loop terms, one finds that~$\Delta'_{con, \mu \nu}(k)$ is the sum of all terms in the skeleton expansion of the connected 4-point function defined above,
\beq
\Delta_{con, \mu \nu}(k) = \Delta'_{con, \mu \nu}(k) + \Delta_{sing, \mu \nu}(k),
\eeq
except for the piece that is connected by a single ghost line $G(k)$,
\beq
\label{missing}
\Delta_{sing}(k) \equiv(G D \Gamma)_\mu(k) \ G(k) \ (\Gamma G D)_\nu(k),
\eeq
which is missing.  Thus, by comparison with (\ref{4pthorizonc}) we obtain the interesting relation
\beq
q - \langle h(0) \rangle = (\Pi + \Delta_{sing})_{\mu \mu}^{aa}(k)|_{k = 0}.
\eeq

	The term ${\Delta_{sing}}_{\mu \mu}^{aa}(k)|_{k = 0}$ is somewhat ambiguous, being of the form $0/0$, but we will determine its value at $k = 0$ by Lorentz invariance.  By Lorentz invariance $(G D \Gamma)_\mu(k) = h(k^2) k_\mu$, and the vector $k_\mu$ has no direction at $k = 0$, so each of the factors in (\ref{missing}) vanishes at $k = 0$.  Thus the missing term in fact vanishes at $k = 0$, and we obtain
\beq
q - \langle h(0) \rangle = \Pi_{\mu \mu}^{aa}(k)|_{k = 0}.
\eeq	
Thus the horizon condition reads
\beq
\label{pihorizon}
\Pi_{\mu \mu}^{aa}(k)|_{k = 0} = 0.
\eeq	
This form of the horizon condition is free of the singular term $\Delta_{sing}(0)$.  

	By Lorentz and color invariance we have
\beq
\Pi_{\mu \nu}^{cd}(k) = [ \ a(k^2) \delta_{\mu \nu} + b(k^2) k_\mu k_\nu \ ] \ \delta^{cd}.
\eeq
Here $a(k^2)$ and $b(k^2)$ are regular at $k = 0$ if $\Pi_{\mu \nu}(k)$ is regular, as appears from its skeleton expansion.  At small~$k$ the second term is negligible compared to the first.  Thus at small $k$ we have
\beq
\Pi_{\mu \nu}(k) \approx a \delta_{\mu \nu},
\eeq
so at small $k$
\beq
k_\mu \Pi_{\mu \nu}(k) k_\nu \approx k^2 \Pi_{\mu \mu}(k)/D,
\eeq
which gives
\beq
\lim_{k \to 0} [ k^2 G(k) ]^{-1} = \Pi_{\mu \mu}(0)/D = 0,
\eeq
by (\ref{PiDSE}) and (\ref{pihorizon}), which establishes (\ref{singghost}), as asserted.

\section{Long range of bose ghost propagator}

	We calculate the bose-ghost propagator in the infrared.  For this purpose we diagonalize the action by writing  the bose ghosts as
\beq
\phi = (U + i V)/ \sqrt 2; \ \ \ \ \ \bar\phi = (U - i V)/ \sqrt 2,
\eeq
where $U$ and $V$ are real fields.  The part of the action that contains $\phi$ and $\bar\phi$ decomposes into a sum of terms that involve $U$ and $V$ separately,
\beq
{\cal L}_{\bar\phi, \phi} = {\cal L}_V + {\cal L}_U + {\cal L}_m,  
\eeq
where
\beq
{\cal L}_V ={1\over 2} \p_\lambda V_\mu^{ab} (D_\lambda V_\mu)^{ab}(y) + i \sqrt 2 \gamma^{1/2} (D_\lambda V_\lambda)^{aa} 
\eeq
\beq
{\cal L}_U = {1 \over 2} \p_\lambda U_\mu^{ab} (D_\lambda U_\mu)^{ab}(y).
\eeq
The mixed term
\beq
{\cal L}_m =
(ig/2) \p_\lambda( f^{abc} V_\mu^{bd} U_\mu^{cd} ) A_\lambda^a
\eeq
vanishes by partial integration when $A$ is transverse, which makes $M(A)$ hermitian.  This mixed term is eliminated in the local formalism by a shift of the Lagrange multiplier field,
\beq
b = b' - (g/2) f^{abc} V_\mu^{bd} U_\mu^{cd}.
\eeq

	The $U$-field does not mix with $A$, and the $U$-$U$ propagator only appears in closed ghost loops, just like the fermi ghosts.  However because of the term $i \sqrt 2 \gamma^{1/2}f^{abc}A_\lambda^b V_\lambda^{ca}$ that is contained in the last term of ${\cal L}_V$, the field $V_\lambda^{ab}$ does mix with the gluon field $A_\lambda^c$ in the sense that the mixed propagator $D_{AV}$, which is given in the Appendix, is non-zero.
	
	We now calculate the long-range part of the $V$-propagator
\beq
D_{VV, \lambda \ \mu}^{\ \ \ \   abcd}(x - y) \equiv 
\langle V_\lambda^{ab}(x) V_\mu^{cd}(y) \rangle.
\eeq 
The $V$-field appears at most quadratically in ${\cal L}_V$, and we obtain a formula for the $V$-propagator, by integrating the $V$-field out of the action $S_V = \int d^Dx \ {\cal L}_V$.  This is done by Gaussian integration after completing the square by shifting the integration variable $V_\lambda^{ab}$ by $i \sqrt 2 g \gamma^{1/2} (M^{-1} A_\lambda)^{ab}$, where
\beq
(M^{-1}A_\lambda)^{ab}(x) = \int d^Dz \ (M^{-1})^{ac}(x, z; A) f^{cdb}A_\lambda^d(z),
\eeq
with the result
\beq
\label{boseprop}
D_{VV, \lambda \ \mu}^{\ \ \ \   abcd}(x - y) = \delta_{\lambda \mu} \delta^{bd} \delta^{ac} G(x-y) + F_{\lambda \ \mu}^{abcd}(x - y),
\eeq
where $\delta^{ac} G(x-y) = \langle (M^{-1})^{ac}(x, y, A) \rangle$ is the fermi-ghost propagator, and
\beq
F_{\lambda \  \mu}^{abcd}(x - y) = - 2 \gamma g^2 \  \langle \ (M^{-1}A_\lambda)^{ab}(x) \ (M^{-1}A_\mu)^{cd}(y) \ \rangle.
\eeq 

	We now show that there is a very long range component to the $F$-term in the $V$ propagator.  By (\ref{kernel}) we may replace $M^{-1}$ by a fermi-ghost pair, so
\beq
F_{\lambda \ \mu}^{abcd}(x - y) = - 2 \gamma g^2 \int d^Dz d^D w \  K_{\lambda \ \mu}^{abcd}(x, z, y, w)  
\eeq
where	
\beqa
K_{\lambda \mu}^{abcd}(x, z, y, w) & \equiv & \langle \ c^a(x) \bar c^e(z) \ f^{egb} A_\lambda^g(z) 
\\ \nonumber 
&& \times \omega_1^{c1}(y) \bar\omega_1^{h1}(w) \  f^{hid} A_\mu^i(w) \ \rangle
\eeqa
is a 6-point correlator.  Here we have arbitrarily chosen one of the equal components of the $\omega$-$\bar\omega$ propagator which are each equal to the propagator of the Faddeev-Popov ghost $\langle \omega_1^{a1}(y) \bar\omega_1^{b1}(w) \rangle = \langle c^a(y) \bar c^b(w) \rangle$.  It may not seem like progress to express the 2-point function $\langle VV \rangle$ in terms of the 6-point function $K$, but in fact the cluster decomposition of $K$ enables us to determine its leading infrared behavior.  The leading term in its cluster decomposition is the product of three connected pieces,
\beqa
K_{3, \lambda \ \mu}^{\ \ abcd}(x, z, y, w) & = & \langle c^a(x) \bar c^e(z) \rangle \  f^{egb} \ \langle A_\lambda^g(z) A_\mu^i(w) \rangle 
\nonumber \\ 
&& \times  f^{hid} \  \langle \omega_1^{c1}(y) \bar\omega_1^{h1}(w) \rangle
\nonumber  \\
& = & G(x-z) f^{aeb} D_{\lambda \mu}(z-w)
\nonumber  \\
&& \times f^{ced} G(w-y) .
\eeqa
The integral over $z$ and $w$ converts this into a convolution which, in momentum space, is the product of propagators,
\beq
\label{F3}
F_{3, \lambda \ \mu}^{\ \ abcd}(k) = - 2 \gamma g^2 \ G(k) f^{aeb} D_{\lambda \mu}(k) f^{ced} G(k).
\eeq
This is highly singular at small $k$. For example if~$G$ and~$D$ were canonical, free, massless propagators (which they are not) the product would be $1/(k^2)^3$.  This is the most infrared singular part of the $V$-ghost propagator, 
\beq
D_{VV, \lambda \ \mu}^{\ \ \ \ abcd}(k) \approx - 2 \gamma g^2 G^2(k) D_{\lambda \mu} (k) f^{aeb} f^{ced},
\eeq
at small $k$.

	We now examine the properties of this asymptotic infrared propagator.  The gluon propagator 
\beq
D_{\lambda \mu}(k) = D(k) P_{\mu \nu}(k)
\eeq
is transverse in Landau gauge, where $P_{\mu \nu}(k) = \delta_{\mu \nu} - k_\mu k_\nu/k^2$ is the transverse projector.  We write the long-range part of the bose ghost propagator as
\beq
D_{VV, \lambda \ \mu}^{\ \ \ \ abcd}(k) \approx P_{\lambda \mu} (k) f^{aeb} f^{ced} D_{VV, lr}(k),
\eeq
which contains transverse and adjoint projectors.  The long-range scalar propagator is given by
\beq
D_{VV,lr} \equiv -2 \gamma g^2 G^2(k) D(k).
\eeq
The combination $g^2 G^2 D$ is a renormalization-group invariant in Landau gauge \cite{Alkofer:2000, vonSmekal:2002}.  This is a consequence of the factorization of the external ghost momenta in the ghost-ghost-gluon vertex, from which it follows that this vertex is finite and does not require renormalization in Landau gauge, so 
\beq
\widetilde Z_1 = Z_g Z_3^{1/2} \widetilde Z_3 = 1.
\eeq
The combination $g^2G^2(k)D(k)$ renormalizes by
$Z_g^2 Z_3 \widetilde Z_3^2 = 1$.
It has been proposed that this combination be used to define a renormalization-group invariant running-coupling in Landau gauge \cite{Alkofer:2000},
\beq
\alpha_s(k) = g^2 (k^2)^3 G^2(k) D(k) / (2\pi).
\eeq

	The infrared behavior of the gluon and ghost propagators is described by infrared exponents~(\ref{infracrit}).  In terms of these we have
\beq
D_{VV, lr}(k) \sim G^2(k) D(k) \sim { 1 \over (k^2)^{3 + \alpha_D + 2 \alpha_G } },
\eeq
which gives for the infrared power law of the bose ghost, 
\beq
\label{Vcriticalexp}
D_{VV, lr}(k) \sim {1 \over (k^2)^{1 + D/2} },
\eeq
in $D$ Euclidean dimensions, as follows from (\ref{exponentrelation}).  

	[We sketch the derivation of equation (\ref{exponentrelation}).  It is the same as in Faddeev-Popov theory, when the horizon condition $[k^2 G(k)]^{-1}|_{k = 0} = 0$ holds \cite{Alkofer:2000, Zwanziger:2001, vonSmekal:2002}, because the DSE of the fermi ghost is the same.  The horizon condition $\Pi_{\mu \nu}(0) = 0$ and the fermi-ghost DSE (\ref{PiDSE}) together imply that the ghost propagator satisfies the equation
\beq
G^{-1}(k) = k_\mu [ \Pi_{\mu \nu}(k) - \Pi_{\mu \nu}(0)] k_\nu,
\eeq
or, by (\ref{pimunu}), 
\beqa
G^{-1}(k) = k_\mu g f [ (D_{\mu \nu} G)(0) -  (D_{\mu \nu} G)(k) 
\nonumber  \\
+  \ \Delta'_{con, \mu \nu}(0) - \Delta'_{con, \mu \nu}(k) ] f g k_\nu.
\eeqa
The terms involving $D_{\mu \nu} G$ represent a loop integral which, for small values of the external momentum $k$, may be evaluated using the asymptotic infrared expressions~(\ref{infracrit}),
\beq
G^{-1}(k) \sim \int d^Dp { k k  \over (p^2)^{1 + \alpha_D} } \Big( {1 \over [(p + k)^2]^{1 + \alpha_G} }
- {1 \over [(p)^2]^{1 + \alpha_G} } \Big).
\eeq
On the left-hand side we have $G^{-1}(k) \sim (k^2)^{1 + \alpha_G}$, whereas on the right-hand side $p$ scales like $k$, so the right-hand side is proportional to $(k^2)^{ -1  - \alpha_D - \alpha_G + D/2}$.  Upon equating like powers of momentum on both sides we obtain (\ref{exponentrelation}).  The correction terms involving $\Delta'_{con}$ do not change this power counting \cite{vonSmekal:2002}.]

	We contrast the exact power law (\ref{Vcriticalexp}) for $D_{VV,lr}$ with the semi-perturbative expansion that starts from the quadratic terms in the local action $\widetilde S$.  (We call it `semi-perturbative' because the non-perturbative horizon condition is imposed after calculating to a given order in perturbation theory.)   Indeed the transverse part of the $V$-propagator, calculated from the quadratic parts of $\widetilde S$, is given by Gribov's expression,
\beq
D_{VV} = D_{AA} = { k^2 \over (k^2)^2 + m^4}.
\eeq	 
This vanishes like $k^2$ at $k = 0$, whereas we the exact result (\ref{Vcriticalexp}) diverges like $1/k^6$ in 4 dimensions.  Thus semi-perturbative theory can be a very bad guide. 

	If the power law of $D_{VV,lr}$, eq~(\ref{Vcriticalexp}), described a potential, it would be over-confining by $1/k^2$.  Indeed in the non-relativistic limit $k \to {\bf k}$, it behaves like $ 1/ {\bf k}^{2 + D} = 1/|{\bf k}|^{3 + S}$, in $S = D -1$ spatial dimensions,  whereas a linearly rising potential in $S$ spatial dimensions behaves like  $1/|{\bf k}|^{1 + S}$.  However we must also consider how the bose ghost $V$ couples to quarks.

\section{Bose ghost as carrier of long-range force}

	The ghost $V$ does not couple directly to quarks, but there is a non-zero q-q-V vertex because the mixed propagator $D_{VA}$ is non-zero.  For example,  the triangle diagram
\beq
\label{triangle}
g^3 \ \lambda^a \gamma_\mu S_{q \bar q} \gamma_\nu \lambda^b \ D_{A  A, \mu\lambda} \ D_{A V, \nu \kappa}^{\ \ \ \  bce} \  f^{acd} \ k_\lambda
\eeq
contributes to the q-q-V vertex. Here $\lambda^a$ are the Gell-Mann matrices, and $g 	f^{acd} \ k_\lambda$ is the tree level V-V-A vertex.  Consequently a $V$-quantum may be exchanged between quarks.  

	The external ghost momentum $k_\lambda$ that appears here is a general feature of the Landau gauge.  It appears at every vertex with an external ghost line, as explained below equation (\ref{ghostDSE}).  This happens for all ghosts including the $V$-ghost because in the local action ${\cal L}_V$, the only vertex in which the $V$-field appears is $(g/2)\p_\lambda V_\mu^{ab} (A_\lambda \times V_\mu)^{ab}$.  Thus, when the $V$ propagator $D_{VV}(k)$ connects two quark lines, there is a factor $k$ at each vertex, so the effective propagator is of the form
\beq
k_\lambda D_{VV,\kappa \nu}(k)  k_\mu \sim {1 \over k^D} = {1 \over k^{S+1}}, 
\eeq 
which rises linearly in position space.  However because there are two Lorentz indices, the coupling is of tensor rather than vector character.
		
	 It would take a non-perturbative, dynamical calculation to determine the q-q-$V$ vertex.  We do not attempt this here.  However we can assert that if this vertex is finite at $k = 0$, after factorization of the external $V$-ghost momentum $k$, then the effective propagator behaves like~$1/k^{1 + S}$.  As an example of what the vertex could be, we may determine its form at small~$k$ if we also assume, for simplicity, that there are no powers of quark momentum at the vertex.  The vertex contains one factor of $k_\mu$, as we have seen, and is contracted with the vector field $V_\nu$.  The long-range propagator term we have found is in the adjoint representation, so this vertex is of the tensor form  $T_{\mu \nu}^a k_\mu$.  Since we assumed that there are no further powers of momentum, it has the Dirac structure $\bar q (A \delta_{\mu \nu} + B \sigma_{\mu \nu}) \lambda^a q \ k_\mu$.  The term $\delta_{\mu \nu} k_\mu = k_\nu$ is longitudinal, and thus orthogonal to the transverse long-range propagator we have just found.  This gives a q-q-$V$ vertex of the form
\beq
\bar q \sigma_{\mu \nu} \lambda^a q \ k_\mu.
\eeq
With this vertex, the net power of the exchange of a $V$ particle between a pair of quarks is given by 
\beq
\bar q \sigma_{\kappa \mu} \lambda^a q \ { k_\kappa k_\lambda \over (k^2)^{1 + D/2} } \  \bar q' \sigma_{\lambda \nu} \lambda^a q'.
\eeq  
In the non-relativistic limit, with spatial dimension $S = D -1$, this reduces to
\beq
\bar q \sigma_i \lambda^a q \ {\delta_{ij} {\bf k}^2 - {\bf k}_k {\bf k}_j \over |{\bf k}|^{3 + S} } \  \bar q' \sigma_j \lambda^a q'.
\eeq
In position space this is a linearly rising, spin-dependent potential.

\section{Conclusion}

	We have found that it is possible to derive exact and unique infrared properties from Faddeev-Popov theory with a cut-off at the Gribov horizon.  The uniqueness contrasts with the situation of the DS equations of the standard Faddeev-Popov action whose solution is not unique because one may cut off the functional integral at any one of the Gribov horizons without changing the DS equations \cite{Zwanziger:2001}, and a one-parameter family of solutions has been found \cite{Fischer:2008}.  It has been proposed to overcome this non-uniqueness by imposing an additional condition $[k^2 G(k)]^{-1}|_{k = 0} = 0$.
		
	 One old lesson learned anew here is that it is dangerous to make approximations when non-perturbative effects are significant.  In particular the ``free" propagator of the bose ghost $V_\mu^{ab}$ that results from the quadratic part of the action $\widetilde S$ is given by $k^2/[(k^2)^2 + m^4]$ which vanishes like $k^2$ in the infrared, whereas we have found that its exact behavior in the infrared is $1/k^6$.  The Gribov pole which occurs at the unphysical value $(k^2)^2 + m^4 = 0$ or $k^2 = \pm i m^2$ could turn out to be an artifact of the free propagator which, as we have  seen, can be highly unreliable.  The long-range propagator of the bose ghost may offer a solution to the problem in Landau gauge of the origin of a long range force between quarks.    	  
	 
	 It is encouraging that the exact properties derived here are consistent with the numerical results of Maas \cite{Maas:2008} that are obtained by fixing to the absolute Landau gauge to the extent practicable.  He has proposed a possible explanation for why numerical results, described in the Introduction, are different in minimal and absolute Landau gauges.  The exact properties we have found disagree with the numerical results in the minimal Landau gauge in $D = 3$ or 4 Euclidean dimensions.  However, as noted in sec.~II, the procedure adopted here, which consists of integrating over the Gribov region, is not the same as a minimal Landau gauge which is algorithm dependent.  This may explain why the properties we find agree rather with the absolute Landau gauge.
			
	The working hypothesis on which the present article is based --- that for purposes of the functional integral the Gribov region $\Omega$ and fundamental modular region $\Lambda$ are equivalent --- does not have as firm a basis as one might wish.  Nevertheless the internal consistency of results derived from this Ansatz is impressive.  Remarkably, the cut-off at the Gribov horizon which is non-local in the gauge field $A_\mu$ turns out to be implementable by a local action that is renormalizable, and moreover the renormalization constants are the same as in Faddeev-Popov theory \cite{Zwanziger:1993}.  This happens because the modifying term is soft, of dimension 2.  The Gribov parameter $\gamma$ does not have an independent subtraction, and the derived condition $[k^2 G(k)]^{-1}|_{k = 0} = 0$ is compatible with multiplicative renormalization.\footnote{This condition agrees with the Kugo-Ojima confinement criterion \cite{Kugo-Ojima}.  The relation between the present approach and the Kugo-Ojima approach, heretofore mysterious has been clarified very recently \cite{Sorella:2009}.}  The ultraviolet structure of the Faddeev-Popov theory, including asymptotic freedom, is preserved.  

	These considerations suggest that cutting off the functional integral at the boundary of the Gribov region~$\Omega$ should extend the range of  applicability of gauge theory toward the infrared, even if ultimately a further refinement will be made in which the cut-off is advanced to the boundary of the fundamental modular region~$\Lambda$.  	
	
	Only infrared properties are modified by a cut-off of Faddeev-Popov functional integral at the Gribov horizon, but here the change is dramatic.   It implies a linearly rising, renormalization-group invariant potential between quarks that is transmitted by the bose ghost, which however has tensor coupling.  Clearly this requires further investigation.
	 
	 Unresolved open questions are unitarity of the physical states, and the relation of confinement to the long-range force found here.  We remain far from a satisfactory understanding the physical states of QCD.   
	 
\bigskip

{\bf Acknowledgements}\\
The author recalls with pleasure stimulating conversations with Laurent Baulieu, Martin Schaden and Alexander Rutenburg.

\appendix

\section{Other exact properties}

	For completeness we note the following relations.  The mixed $V$-$A$ propagator is given by
\beq
\langle V_\lambda^{ab}(x) A_\mu^c(y) \rangle = - i g (2 \gamma)^{1/2} \langle (M^{-1}A_\lambda)^{ab}(x) A_\lambda^c(y) \rangle,
\eeq	
and the $U$ propagator by
\beqa
\langle U_\lambda^{ab}(x) U_\mu^{cd}(y) \rangle  & = & \langle \delta_{\lambda \mu} \delta^{bd}(M^{-1})^{ac}(x, y, A) \rangle
\nonumber  \\
& = & \delta_{\lambda \mu} \delta^{bd}G^{ac}(x-y).
\eeqa
where $G^{ac}(x-y)$ is the propagator of the Faddeev-Popov ghost.  The auxiliary fermi ghosts $\omega$ and~$\bar\omega$ have the same propagator as $c$-$\bar c$, and $U$-$U$,
\beq
\langle \omega_\lambda^{ab}(x) \bar\omega_\mu^{cd}(y) \rangle = \delta_{\lambda \mu} \delta^{bd}G^{ac}(x-y).
\eeq
One may also show in the infrared limit
\beq
D_{AA}(k) D_{AV}(k) - D_{AV}^2(k) = O(1).
\eeq

\appendix

%%%%%%%%%%%%%%%%%%%%%%%%%%%%%%%%%%%%%%%%

%%%%%%%%%%%%%%%%%%%%%%%%%%%%%%%%%%%%%%%%%%%%%%%%%%%	

\end{document}